\documentclass[]{spie}  

\usepackage{amsmath,amsfonts,amssymb}
\usepackage{graphicx}
\usepackage[colorlinks=true, allcolors=blue]{hyperref}
\usepackage{wrapfig}
\usepackage{lscape}
\usepackage{rotating}
\usepackage{epstopdf}
\usepackage{subcaption}
\usepackage{setspace}
\usepackage{lineno}
\usepackage{multirow}
\usepackage{float}

\title{Tiny Observatory for Telescope Optimization (TOTO): testing algorithms for autonomous on-orbit alignment for space-based telescope systems}

\author[a,b]{Solvay A. Blomquist}
\author[a]{Sanchit Sabhlok}
\author[a,b]{Hyukmo Kang}
\author[a]{Maggie Y. Kautz}
\author[a]{Simran Agarwal}
\author[a,b]{Heejoo Choi}
\author[a,b]{Alexandra Kupersmith}
\author[a]{Adam Schilperoort}
\author[a]{Kelsey L. Miller}
\author[a]{Kyle J. Van Gorkom}
\author[a]{Corey Fucetola}
\author[a]{Patrick Ingraham}
\author[a]{Ewan S. Douglas}
\author[a,b]{Daewook Kim}
\affil[a]{Steward Observatory - The University of Arizona \linebreak 933 N Cherry Ave, Tucson, AZ 85721 \linebreak}
\affil[b]{Wyant College of Optical Sciences - The University of Arizona \linebreak 1630 E University Blvd, Tucson, AZ 85721}

\authorinfo{Further author information: (Send correspondence S.B.)\\S.B. E-mail: sablomquist@arizona.edu}

\begin{document} 
\maketitle

\begin{abstract}
The Tiny Observatory for Telescope Optimization (TOTO) is an optical testbed designed to evaluate the efficacy of autonomously driven alignment algorithms for space-based telescope systems. For space-based missions, active control of the telescope alignment on-orbit offers potential to relax passive alignment requirements and reduce on-ground verification activities. TOTO is used to evaluate and verify simulation work of two primary alignment algorithms, Stochastic Parallel Gradient Descent (SPGD) and focus-diverse phase retrieval (FDPR). Previous simulation work has confirmed that by using SPGD for coarse alignment followed by focus-diverse phase retrieval for fine alignment, we can reach diffraction-limited performance on-orbit. This paper presents the results of the autonomous alignment algorithm of a Cassegrain telescope using TOTO. We report the current status of TOTO as well as preliminary results from SPGD and phase retrieval on the testbed using monochromatic light source to simulate an on-axis point source.

\end{abstract}


\section{Introduction}

For current and future space-based observatories, it is advantageous to implement wavefront sensing and control algorithms to maintain diffraction limited imaging for science acquisition. To test the accuracy and efficiency of these algorithms in controlling the quality of point spread function (PSF), optical testbeds are built to mimic the physics of the telescope under study and monitor system performance while undergoing alignment and control. The Tiny Observatory for Telescope Optimization (TOTO) is an optical testbed designed to test closed-loop correction methods on-orbit for space-borne telescopes \cite{kang2025design}. The studies presented in this paper discuss the process of aligning TOTO as well as the strategies used to verify simulations results from alignment processes such as focus-diverse phase retrieval (FDPR) and Stochastic Parallel Gradient Descent (SPGD) in a controlled lab environment \cite{blomquist2025stages, derby2025fine, derby2023integrated, sanch_talk}.

\newpage

\section{Coarse On-Orbit Alignment and Testing} 
\subsection{TOTO Configuration}
The design of TOTO is addressed in Kang et al. 2025 with the main capabilities of testing alignment on TOTO building upon an existing Ritchey-Chretien telescope design as well as an ALPAO 97-15 13.5 mm deformable mirror (DM)\cite{ashcraft2021versatile, kang2025design}. The primary mirror (M1) is currently mounted to a stationary stage while the secondary mirror (M2) is mounted to a fully motorized translation stage. The tip and tilt for both mirrors are fully motorized using piezoelectric inertia motors. The current design of TOTO has the DM conjugate to the primary mirror to simulate low-order aberrations of the primary mirror that will be corrected through bending modes \cite{blomquist2023analysis}.  

\begin{figure}[hbt!]
    \centering
    \includegraphics[width=1\linewidth]{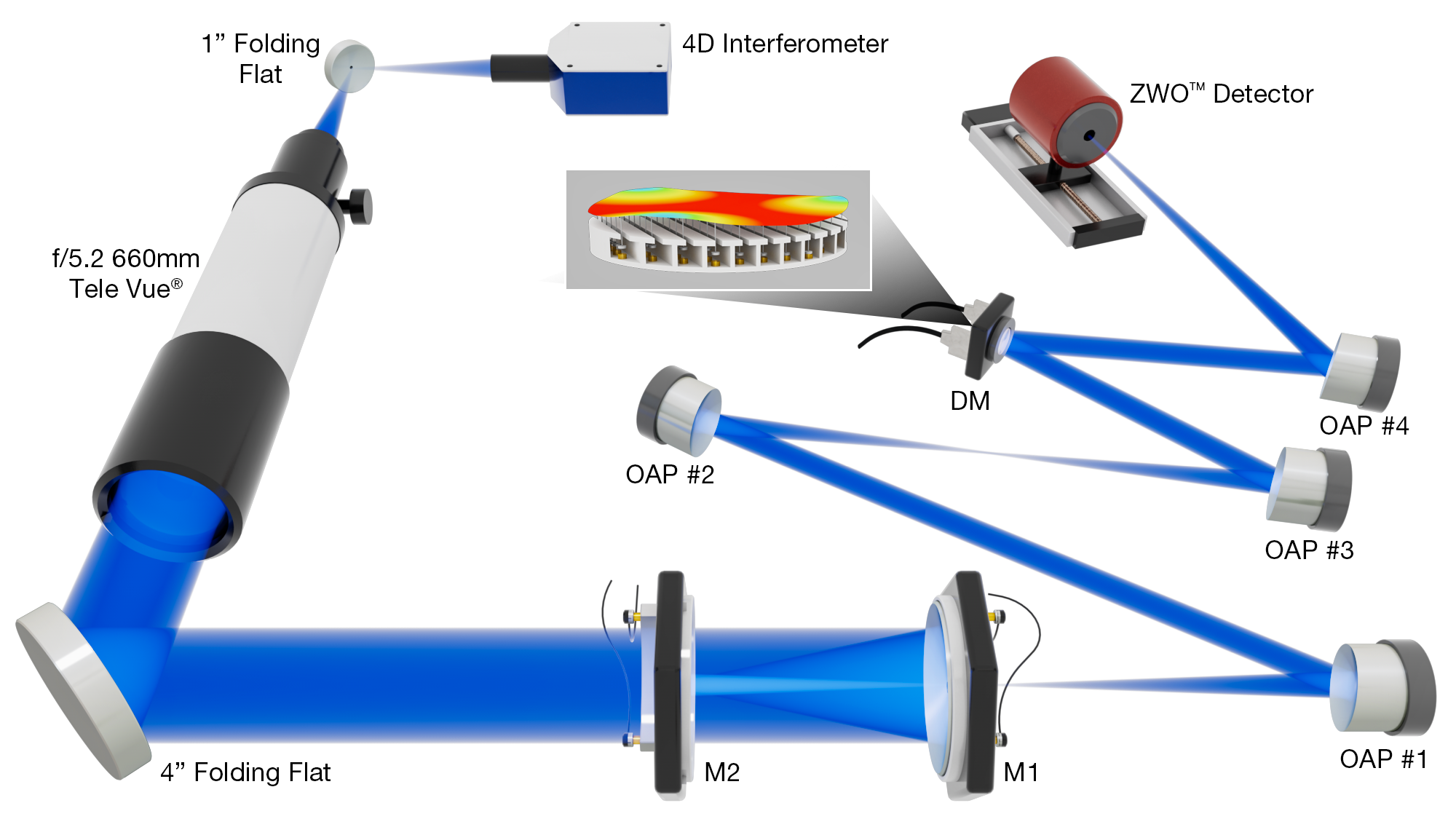}
    \caption{Layout of all components on TOTO testbed. Graphic  created by Steven Burrows.}
    \label{fig:toto_layout}
\end{figure}

From the original design in Kang et al. 2025, we have made changes to the optical design to increase space on the optical bench without affecting the optical prescription of the design. Figure \ref{fig:toto_layout} shows the optical layout of TOTO. Due to limited space on the optical bench and to correct residual tip/tilt from the 4D interferometer, we added a 1" fold flat immediately following the objective lens of the interferometer. To achieve a 3.5" beam from our point source to the primary mirror, we use a Tele Vue-NP127is Nagler-Petzval $f/5.2$ refractor to propagate an expanded and collimated beam out to a 4" folding flat mirror. The 4" folding flat is mounted to a motorized rotation stage to perform field selection for future tests. For all tests conducted in this paper, we use a 4D PhaseCam 6000 interferometer to simulate an on-axis point source star. 

For the results presented in this paper, we are only considering the misalignment and correction for available rigid-body degrees of freedom on M2 and deformable mirror. For follow-up work, we will be incorporating tests using both M1 and M2 in coarse alignment procedures. After aligning TOTO and cleaning up residual misalignment from coarse alignment, we had a suitable PSF to then measure the sensitivity of the system to M2 motion.

\subsection{M2 Sensitivity}

To inform the SPGD algorithm of gain parameters to apply when moving M2, a sensitivity analysis was conducted on TOTO in comparison to simulation data from Zemax OpticStudio. First, we started by moving M2 in the three translation degrees of freedom, $[D_x, D_y, D_z]$ in Zemax and measured the response of the RMS spot size with respect to the respective motion in translation. 

\begin{figure}[hbt!]
    \centering
    \begin{minipage}{\textwidth}
        \centering
        \begin{subfigure}{\textwidth}
            \centering
            \includegraphics[width=0.95\textwidth]{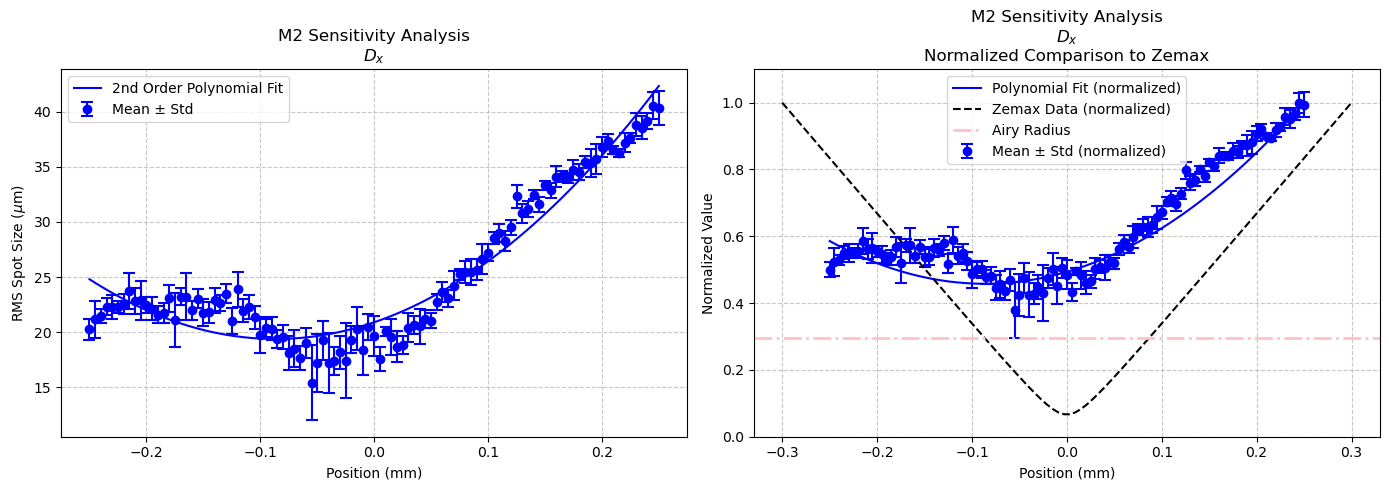}
            \caption{Sensitivity data from TOTO measuring RMS spot size as a function of translating M2 in $D_x$.}
        \end{subfigure}
    \end{minipage}

    \begin{minipage}{\textwidth}
        \centering
        \begin{subfigure}{\textwidth}
            \centering
            \includegraphics[width=0.95\textwidth]{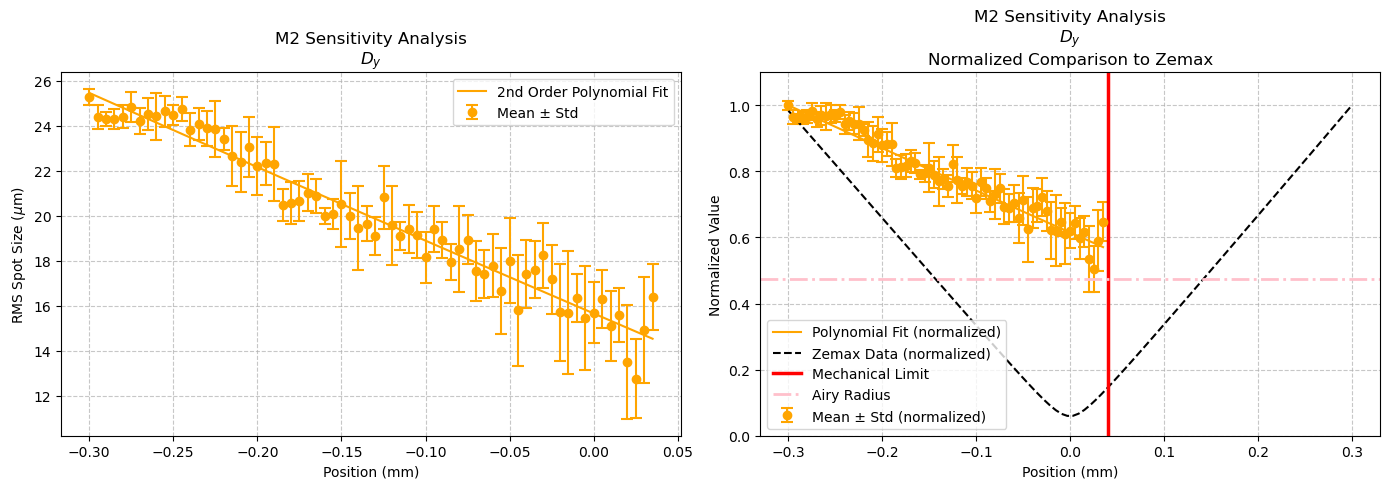}
            \caption{Sensitivity data from TOTO measuring RMS spot size as a function of translating M2 in $D_y$. Note that the raw data is truncated in $D_y$ due to the nominal position of M2 in the y-axis being close to the limit of the range of motion on the translation stage. Note that Zemax can sense RMS spot size below the system Airy ring size, which cannot be physically sensed on the testbed.}
        \end{subfigure}
    \end{minipage}

    \begin{subfigure}{\textwidth}
        \centering
        \includegraphics[width=0.95\textwidth]{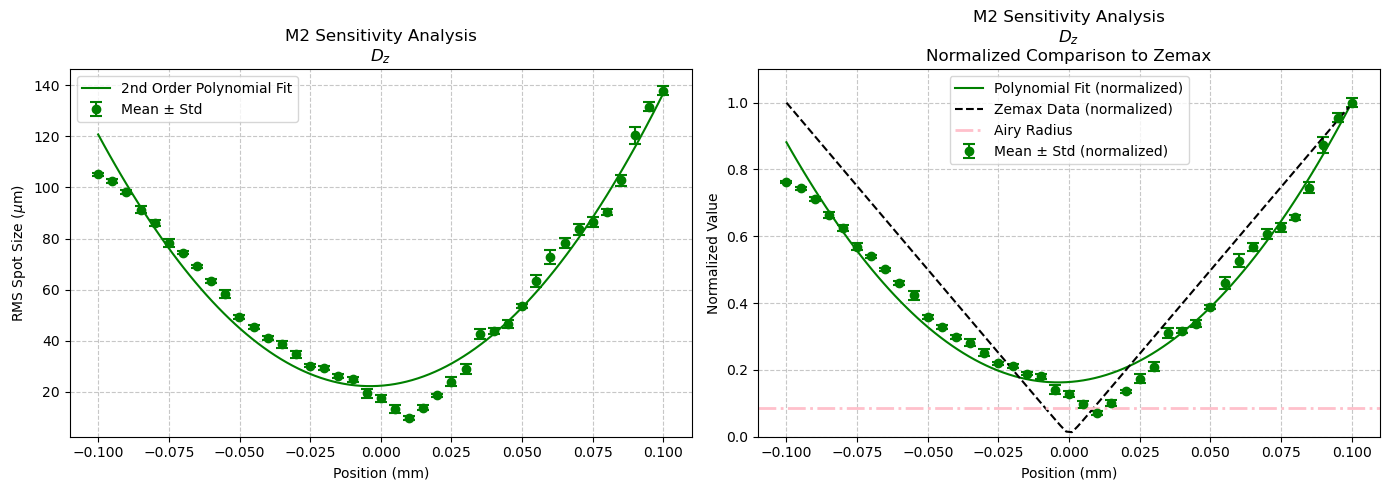}
        \caption{Sensitivity data from TOTO measuring RMS spot size as a function of translating M2 in $D_z$}
    \end{subfigure}

    \caption{Plots showing RMS spot size as a function of moving M2 in the three available translational degrees of freedom, [$D_x, D_y, Dz$]. The plots on the right of all three show the normalized data in comparison to Zemax normalized data.}
    \label{fig:testbed_sens_plots}
\end{figure}

For comparing RMS spot size from Zemax model to the as-built testbed system, we use a similar method of calculating spot size in respect to the spot centroid. The current algorithm used on the testbed assumes the position of the pixel with the highest flux as the centroid. It then applies a threshold mask to the surrounding pixels whose flux is greater than $10\%$ of the maximum flux. We then perform a flux weighted calculation to get an accurate RMS spot size comparable to the results from Zemax. The results of that calculation in response to moving M2 in [$D_x, D_y, D_z$] are shown in Figure \ref{fig:testbed_sens_plots}.

\begin{figure}[hbt!]
    \centering
    \includegraphics[width=0.95\linewidth]{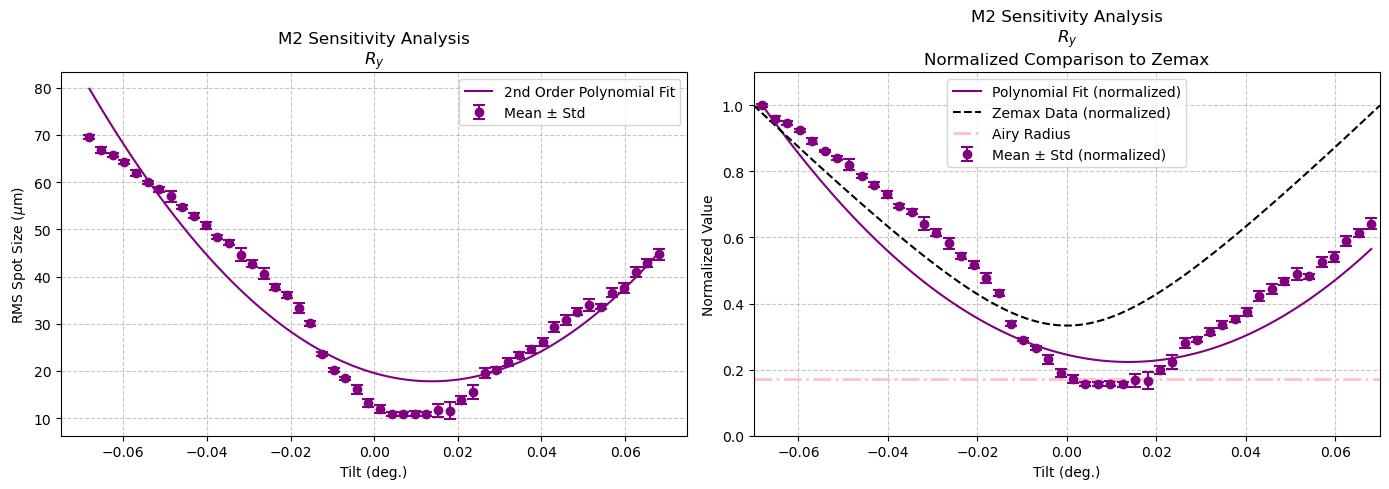}
    \label{fig:enter-label}
    \caption{Sensitivity data from TOTO measuring RMS spot size as a function of translating M2 in $R_y$. Discrepancies between model and testbed data may be due to coupling between de-focus and translation of the optic when mechanically tilting the optic.}
    \label{fig:testbed_sens_ry}
\end{figure}

When measuring the sensitivity of the system in response to moving M2 in Tilt X and Tip Y ([$R_x, R_y$]), there was mechanical error in moving the Tilt X axis on M2. For the tests demonstrated in this paper, we will be using only Tip Y ($R_y$). Tilt X will be used in follow-up experiments on TOTO once the actuator is functional. Figure \ref{fig:testbed_sens_ry} shows the response of RMS spot size as a function of moving M2 in $R_y$ on TOTO in comparison to Zemax.

From Figure \ref{fig:testbed_sens_ry} and \ref{fig:testbed_sens_plots} it is apparent that $D_z$ and $R_y$ are the most sensitive to misalignment, while $D_x$ and $D_y$ may require more misalignment to see a meaningful convergence with SPGD. Based on these results, we can gauge a reasonable starting misalignment position for testing how well we can converge with SPGD given current hardware limitations.

\newpage

\newpage

\section{Automated Alignment Procedure}

\begin{figure}[hbt!]
    \centering
    \includegraphics[width=0.5\linewidth]{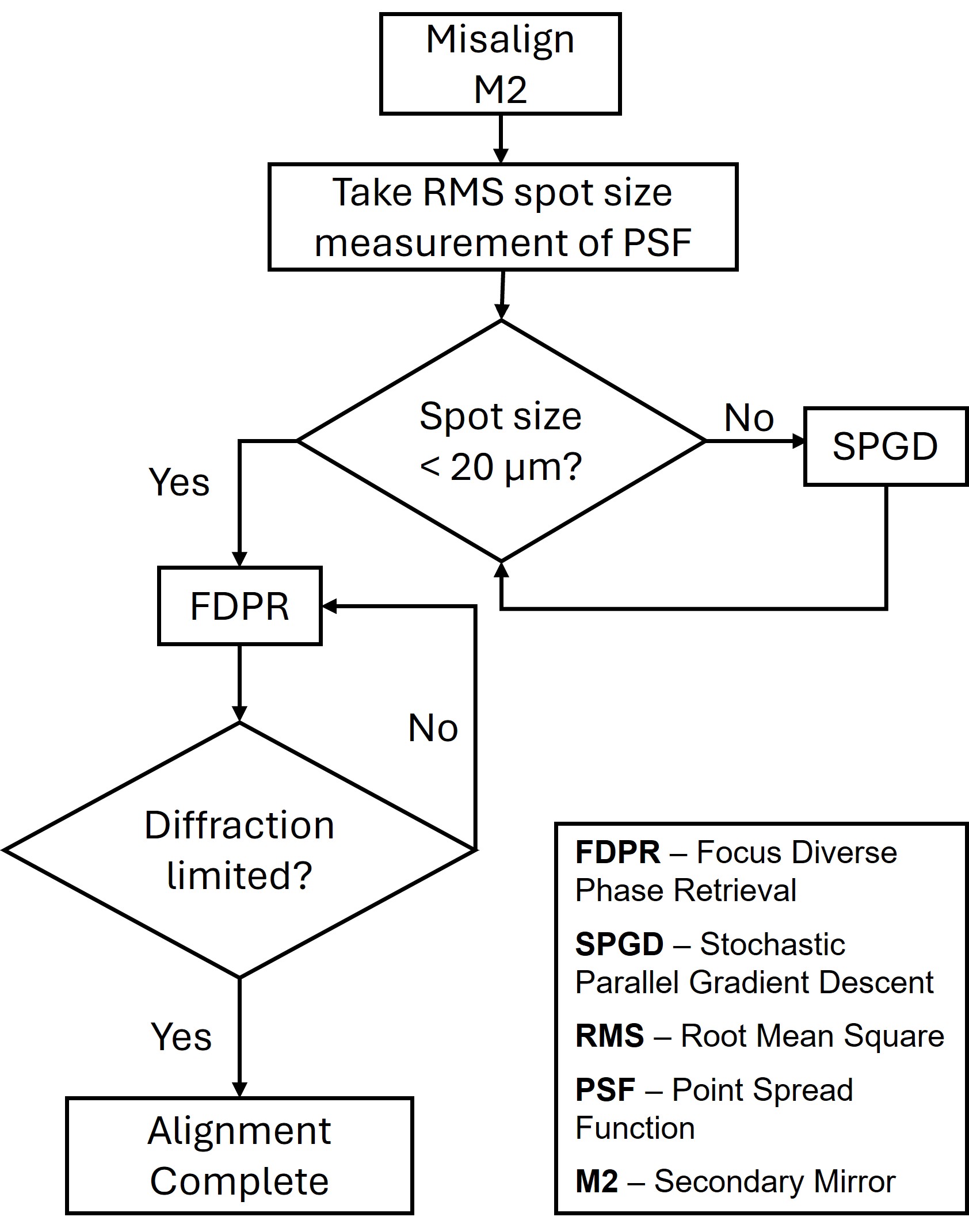}
    \caption{Flowchart of alignment procedure on TOTO.}
    \label{fig:toto_flow}
\end{figure}

For validating alignment procedures outlined in simulation done in previous works, we employ the following flow chart, shown in Figure \ref{fig:toto_flow}, as a way to assess spot size and employ different alignment techniques. The process begins by misaligning M2 by some amount, the misalignment values used in later sections are chosen such that each process can be thoroughly tested. For SPGD and FDPR, each process is iterated through in a loop until their respective criterion is met. When employing FDPR, we continuously sense and apply error at the DM until we have reached the diffraction limited regime for our system.

\section{Experimental Demonstrations}

\subsection{Stochastic Parallel Gradient Descent (SPGD)}

\begin{figure}[hbt!]
    \centering
    \includegraphics[width=0.8\linewidth]{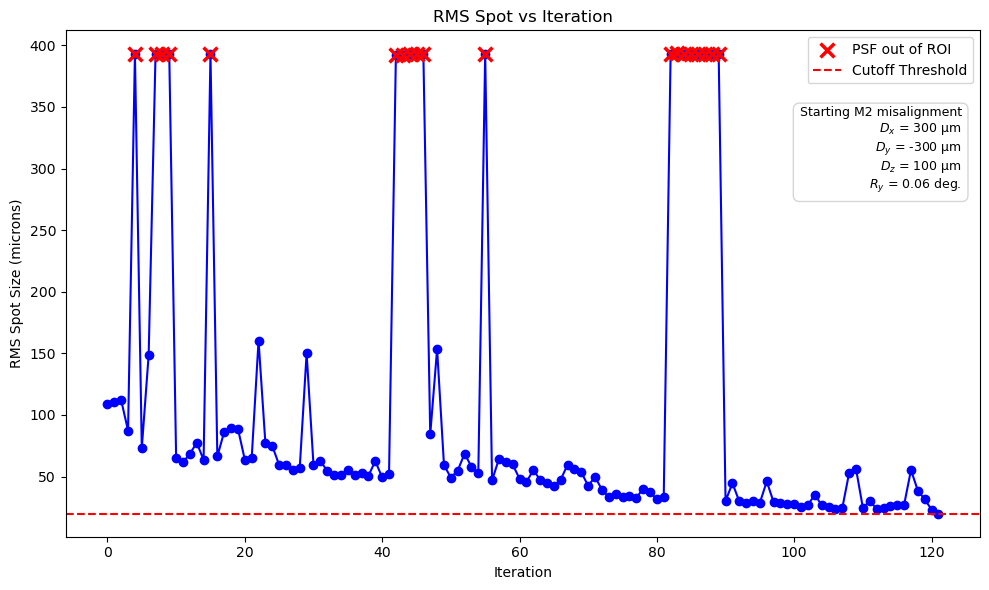}
    \caption{SPGD on all available degrees of freedom misaligned on TOTO. The cutoff threshold of 20 $\mu$m was based on the sensitivity data as a feasible goal which all degrees of freedom were able to reach on the testbed.}
    \label{fig:spgd_all_dof}
\end{figure}

For testing SPGD on TOTO, we first started by individually misaligning each translational degree of freedom and running the algorithm on that specific degree of freedom only. The algorithm used on the testbed is comparable to the one described in Blomquist et al. 2025 \cite{blomquist2025stages}. The gain parameters used in simulation versus on the testbed slightly differ due to the M2 sensitivity analysis shown in Section 2.1 of these proceedings. The simulations done prior were done with perfect systems with no noise nor positional backlash seen in a real system. When measuring the stability of the RMS spot size after SPGD has completed, we let the system continuously measure the spot size for several iterations after the cutoff threshold has been met and demonstrated the system to remain stable after it had reached cutoff criterion. Following these individual tests, we then employed SPGD on all available degrees of freedom, [$D_x, D_y, D_z, R_y$]. RMS spot size as a function of iteration of the SPGD algorithm is shown in Figure \ref{fig:spgd_all_dof}.

After running SPGD on TOTO, we confirmed its convergence with all available degrees of freedom misaligned, the next step is to confirm focus-diverse phase retrieval (FDPR) on the testbed before testing the two processes together.

\subsection{Focus-Diverse Phase Retrieval (FDPR)}

After SPGD is complete, the next stage of alignment is focus-diverse phase retrieval (FDPR). On TOTO currently, there are two ways in which focus diversity can be applied to the system. The first method is through applying defocus using the ALPAO deformable mirror \cite{vanGorkom2021:DMCharacterization, vanGorkom2022:Scoob2}. The second method is applying defocus by moving the detector, mounted on a translation stage. We use the first method for open and closed loop correction of the PSF. We briefly summarize the method here. 

\begin{figure}[hbt!]
    \centering
    \includegraphics[width=0.9\linewidth]{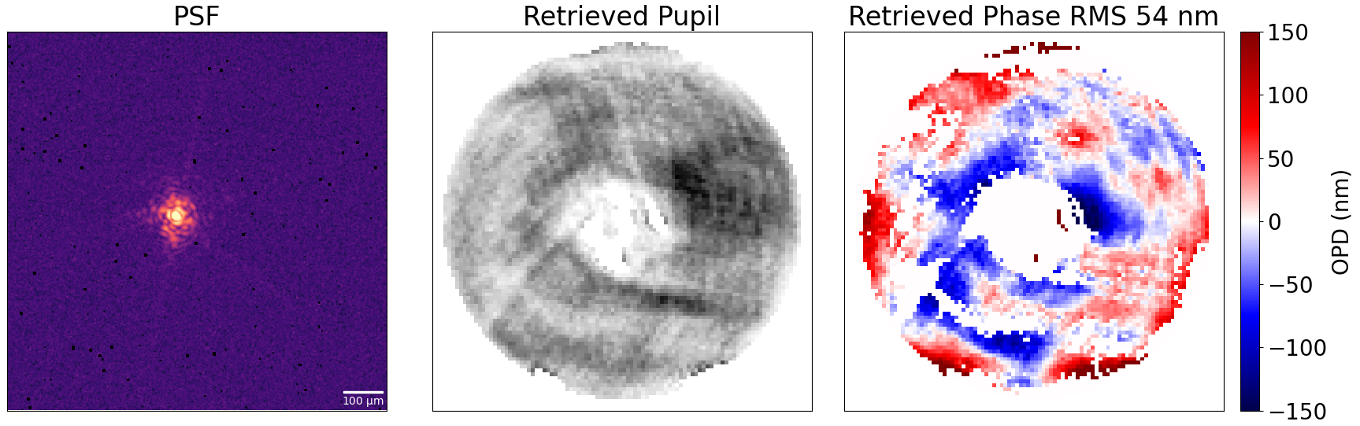}
    \caption{From left to right, the aligned point spread function (PSF), retrieved pupil mask and phase after running FDPR using ALPAO 97 actuator deformable mirror. }
    \label{fig:phase_retriev_before}
\end{figure}

The algorithm sets a defocused Zernike polynomial on the DM, normalized to an RMS value of 1 $\mathrm{\mu m}$. We use 4 values of defocus (-0.25, -0.1, 0.1, 0.25) $\mathrm{\mu} m$ on the DM for each iteration of phase retrieval, both in open and closed loop. All defocus images are used to construct a forward and reverse gradient model of the PSF \cite{Jurling2014:AlgoDiff, Thurman2009:FDPR}, which are simultaneously minimized to obtain the best incident wavefront phase and amplitude. Figure \ref{fig:phase_retriev_before} shows the result from a single iteration of FDPR. The FDPR algorithm in closed loop is generally run for 10 iterations, even if reasonable convergence is found in 3-5 iterations. 

\newpage

\subsection{SPGD \& FDPR}

\begin{figure}[htbp]
  \centering
  
  \begin{subfigure}[b]{0.38\textwidth}
    \centering
    \includegraphics[width=\textwidth]{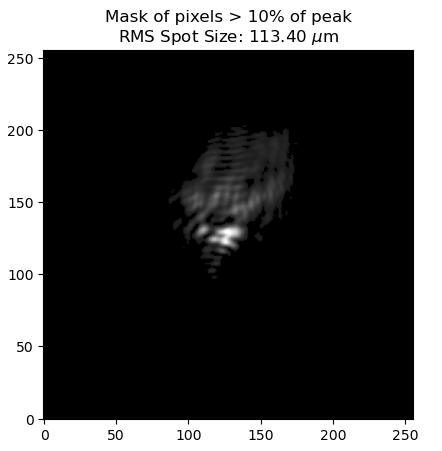}
    \caption{Detected PSF before SPGD.}
    \label{fig:left_plot}
  \end{subfigure}
  \begin{subfigure}[b]{0.38\textwidth}
    \centering
    \includegraphics[width=\textwidth]{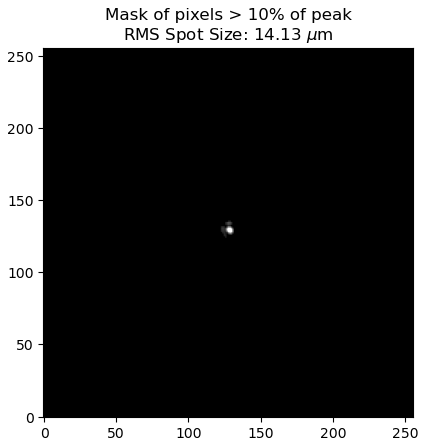}
    \caption{PSF after SPGD.}
    \label{fig:right_plot}
  \end{subfigure}

  \caption{Detected PSF shape and RMS spot size before and after SPGD. Note that the detected PSF size after running SPGD is well below the 20 $\mu$m RMS spot size cut off.}
  \label{fig:psf_comp}
\end{figure}

After testing and validating both algorithms individually, the final step for testing the entire algorithm in Figure \ref{fig:toto_flow} is to test a seamless transition from SPGD to FDPR. As shown in Figure \ref{fig:toto_flow}, the first step of alignment is for SPGD to go from some misaligned PSF to near diffraction limited for phase retrieval to sense the remaining phase error. Figure \ref{fig:psf_comp} shows the PSF before and after running SPGD on TOTO. 

The corresponding convergence map is shown below in Figure \ref{fig:spgd_on_toto}.

 \begin{figure}[hbt!]
    \centering
    \includegraphics[width=0.8\linewidth]{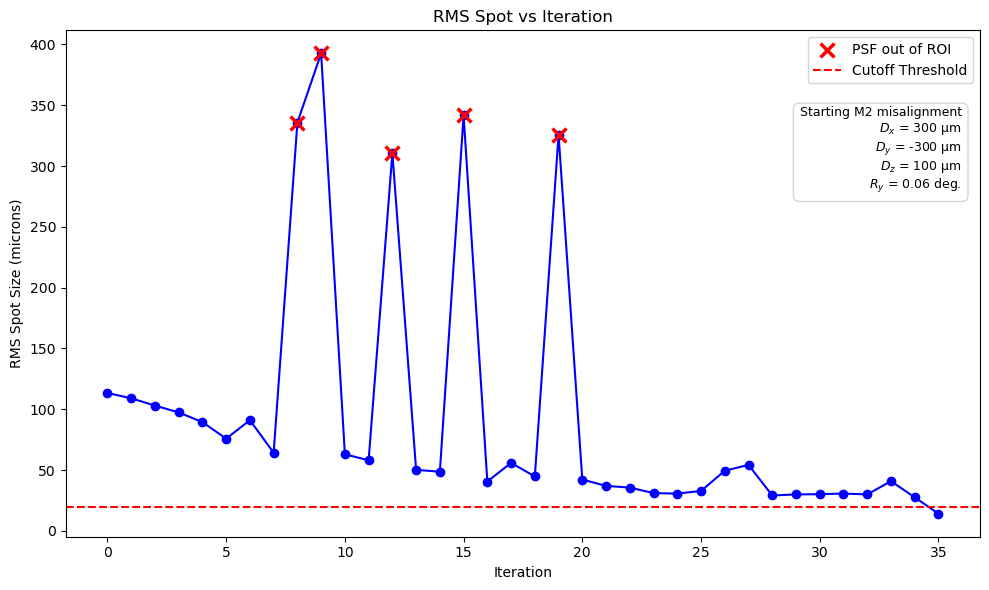}
    \caption{Convergence map of SPGD from a starting PSF shown in Figure \ref{fig:left_plot} to the final state below the cut off threshold with the corresponding PSF shown in \ref{fig:right_plot}.}
    \label{fig:spgd_on_toto}
\end{figure}

\begin{figure}[hbt!]
    \centering
    \includegraphics[width=0.9\linewidth]{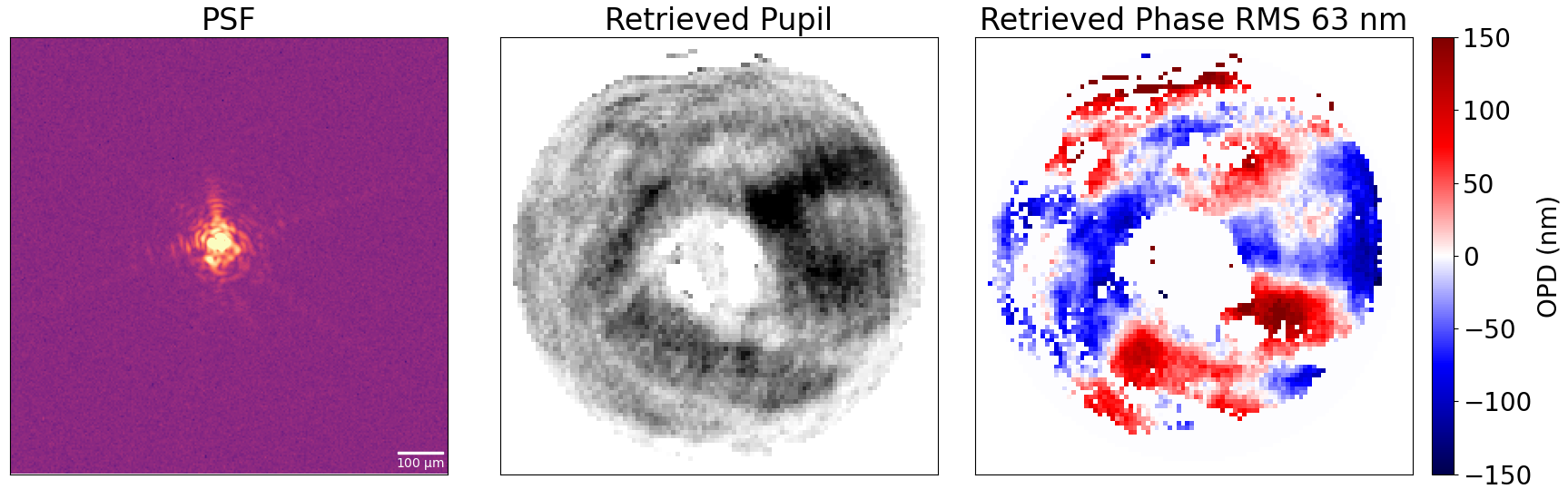}
    \caption{From left to right, the log-scaled PSF after SPGD, retrieved corresponding pupil mask and phase after running FDPR on the post-SPGD PSF.}
    \label{fig:phase_retriev_after_spgd}
\end{figure}

Finally, we have demonstrated that we can sense remaining error not corrected via SPGD using FDPR by applying defocus to the DM. The final sensed phase error map and corresponding pupil map is shown in Figure \ref{fig:phase_retriev_after_spgd}.

\section{Conclusions}

These preliminary results from TOTO demonstrate that we can effectively use the alignment algorithms previously tested in simulation for autonomous alignment. These tests also show the potential for these alignment algorithms to be used in closed-loop correction until we can fully reach diffraction limited performance as the final aligned state of the testbed.

Future works include demonstrating these alignment capabilities for all degrees of freedom for both the primary and secondary mirrors. Also, evaluating sensitivity of the system in the context of Zernike coefficients to help further inform the phase retrieval algorithm. For continued testing moving towards a realistic environment, we will shift from a monochromatic light source to a broadband source and also use the rotational capabilities of the 4" flat mirror to simulate guide stars both on- and off- axis within the telescope field of view.

\section*{ACKNOWLEDGMENTS}

Portions of this research were supported by funding from the Technology Research Initiative Fund (TRIF) of the Arizona Board of Regents and by generous philanthropic donations to the Steward Observatory of the College of Science at the University of Arizona.

\bibliography{report} 
\bibliographystyle{spiebib} 

\end{document}